\documentclass[a4paper,conference]{IEEEtran} 
\IEEEoverridecommandlockouts
\usepackage{cite}
\usepackage{amsmath,amssymb,amsfonts}

\usepackage{graphicx}
\usepackage{textcomp}
\usepackage{xcolor}

\usepackage{times}
\usepackage{epsfig}
\usepackage{graphicx}
\usepackage{amsmath}
\usepackage{amssymb}
\usepackage{algorithm}
\usepackage{algpseudocode}

\usepackage{indentfirst}
\usepackage{multirow}
\usepackage{rotating}
\usepackage{array}
\usepackage{makecell}
\usepackage{tabularx}

\newcolumntype{C}[1]{>{\centering\arraybackslash}m{#1}}
\newcolumntype{Y}{>{\centering\arraybackslash}X}

\usepackage{bm}
\usepackage{amssymb}
\usepackage{pifont}

\usepackage{xcolor}
\usepackage{soul}
\usepackage{float}
\usepackage{cleveref}
\usepackage{layouts}
\usepackage{subcaption}

\usepackage[caption=false, font=normalsize, labelfont=sf, textfont=sf]{subfig}

\def\BibTeX{{\rm B\kern-.05em{\sc i\kern-.025em b}\kern-.08em
    T\kern-.1667em\lower.7ex\hbox{E}\kern-.125emX}}
    
\begin{document}

\setlength{\textfloatsep}{4pt}
\setlength{\intextsep}{4pt}
\setlength{\floatsep}{4pt}

\title{Poisoning Attacks on Federated Learning-based Wireless Traffic Prediction
\thanks{ISBN 978-3-903176-63-8 © 2024 IFIP}
}

\author{
		\IEEEauthorblockN{Zifan Zhang$\IEEEauthorrefmark{1}$,
		Minghong Fang$\IEEEauthorrefmark{2}$,
            Jiayuan Huang$\IEEEauthorrefmark{1}$,
            Yuchen Liu$\IEEEauthorrefmark{1}$
 }
		\IEEEauthorblockA{
		$\IEEEauthorrefmark{1}$North Carolina State University, USA,
		$\IEEEauthorrefmark{2}$University of Louisville, USA
		}
	}

\maketitle

\begin{abstract}
Federated Learning (FL) offers a distributed framework to train a global control model across multiple base stations without compromising the privacy of their local network data. This makes it ideal for applications like wireless traffic prediction (WTP), which plays a crucial role in optimizing network resources, enabling proactive traffic flow management, and enhancing the reliability of downstream communication-aided applications, such as IoT devices, autonomous vehicles, and industrial automation systems. Despite its promise, the security aspects of FL-based distributed wireless systems, particularly in regression-based WTP problems, remain inadequately investigated. In this paper, we introduce a novel fake traffic injection (FTI) attack, designed to undermine the FL-based WTP system by injecting fabricated traffic distributions with minimal knowledge. We further propose a defense mechanism, termed global-local inconsistency detection (GLID), which strategically removes abnormal model parameters that deviate beyond a specific percentile range estimated through statistical methods in each dimension.
Extensive experimental evaluations, performed on real-world wireless traffic datasets, demonstrate that both our attack and defense strategies significantly outperform existing baselines.

\end{abstract}


\section{Introduction}
\label{sec:intro}

Federated learning (FL) represents an evolving paradigm in distributed machine learning techniques, allowing a unified model to be trained across numerous devices containing local data samples, all without the need to transmit these samples to a central server. This innovative framework empowers training on diverse datasets characterized by heterogeneous distributions, offering substantial advantages in the current landscape of big data. In practical applications, FL has found widespread use in addressing real-world challenges, particularly in environments dealing with sensitive or personal data, including the Internet of Things (IoT)~\cite{khan2021federated, nguyen2021federated}, edge computing~\cite{abreha2022federated}, and health informatics~\cite{xu2021federated,nguyen2022federated}. 

In the realm of wireless networks, FL leverages its distributed nature to facilitate multiple network services, including wireless traffic prediction (WTP).
With the exponential growth in the number of connected devices and the ever-increasing demand for data-intensive applications like streaming, online gaming, and IoT services, predicting wireless traffic accurately becomes vital for ensuring network reliability and efficiency. By forecasting network load on a temporal basis, service providers can dynamically allocate resources, reducing the risk of congestion and ensuring a high Quality of Service (QoS) for users~\cite{wen2020assisting, nie2017network, sone2020wireless}. Furthermore, accurate traffic predictions enable operators to strategically plan network expansions and efficiently upgrade infrastructure, resulting in cost savings and enhanced network performance. Particularly, in the era of 5G and beyond, where technologies like network slicing and edge computing play crucial roles, WTP becomes essential for optimizing these advancements, which not only enhances user experience but also facilitates the provision of innovative services that demand high bandwidth and low latency.
To implement WTP, while centralized methods exist~\cite{qiu2018spatio, xu2017high}, FL-based solution stands out by utilizing training data distributed across diverse edge nodes. This approach enhances the generation of precise and timely predictions concerning network traffic~\cite{zhang2021dual}. 
Despite FL's potential in accuracy, efficiency, and privacy preservation, its integration into WTP is not devoid of challenges. Notably, Byzantine attacks, particularly model poisoning attacks, pose significant threats to the effectiveness and trustworthiness of FL-based WTP systems~\cite{zheng2022poisoning}. 

In a model poisoning attack, malicious network entities introduce adversarial modifications to the model parameters during training process of WTP. This tampering results in a compromised global model when aggregated at the central network controller, subsequently producing incorrect traffic predictions. Such inaccuracies lead to the risk of network inefficiencies and even severe service disruptions, especially in real-time applications like autonomous driving systems. In more extreme scenarios, these attacks may serve as gateways to further malicious network intrusions, instigating broader security and privacy concerns as illustrated in~\cite{joshi2015review, fan2019research}. The grave implications of model poisoning attacks underscore the pressing need for robust security measures to ensure the integrity, reliability, and resilience of FL-based WTP systems against Byzantine failures, thereby safeguarding the overarching network infrastructure and the services reliant on it. While most existing FL algorithms and their associated security strategies are typically assessed within the context of \textit{classification} problems~\cite{fang2020local,shejwalkar2021manipulating}, scant attention has been paid to the \textit{regression} problems, as observed in examined WTP scenarios, introducing distinct challenges related to data distribution, model complexity, and evaluation metrics. The distinction between data manipulation strategies in regression and classification problems, as well as their detection methodologies, underscores the nuanced challenges in safeguarding machine learning models against attacks. For instance, in regression-based WTP problems, attackers typically target the model's continuous output by altering the distribution or magnitude of input time-series data, with the goal of steering predictions in a specific direction. This differs from classification tasks, where the manipulation revolves around modifying input features to induce misclassification without noticeably changing the input's appearance to human observers.

To bridge this gap, we make the first attempt to introduce a novel attack centered on injecting fake base station (BS) traffic into wireless networks. Existing model poisoning attacks have predominantly depended on additional access knowledge and direct intrusions on BSs~\cite{zheng2022poisoning, fang2020local, xie2020fall}. However, in practical cellular network systems, BSs have exhibited a commendable level of resilience against attacks, making the extraction of training data from them a challenging endeavor. In contrast, the cost of deploying fake BSs that mimic their behaviors is comparatively lower than the resources required for compromising authentic ones~\cite{cao2022mpaf}. This assumption asserts that these compromised BSs lack insight into the training data and only have access to the initial and current global models, aligning with the practical settings studied in~\cite{cao2022mpaf}. Importantly, other information, such as data aggregation rules and model parameters from benign BSs, remains inaccessible to these compromised BSs. Within the FL framework, the global model is aggregated based on the model parameters of BSs in each iterative round, encompassing both benign and fake BSs. Consequently, our threat model envisions a minimum-knowledge scenario for an adversary. To this end, we propose Fake Traffic Injection (FTI), a methodology designed to create undetectable fake BSs with minimal prior knowledge, where each fake BS employs both its initial model and current global information to determine the optimizing trajectory of the FL process on WTP. 
These malicious participants aim to subtly align the global model towards an outcome that undermines the integrity and reliability of the data learning process.
Numerous numerical experiments are conducted to validate that our FTI demonstrates efficacy across various state-of-the-art aggregation rules, outperforming other model poisoning attacks in terms of vulnerability impacts.

On the contrary, we propose an innovative defensive strategy known as Global-Local Inconsistency Detection (GLID), aimed at neutralizing the effects of model poisoning attacks on WTP. This defense scheme involves strategically removing abnormal model parameters that deviate beyond a specific percentile range estimated through statistical methods in each dimension. Such an adaptive approach allows us to trim varying numbers of malicious model parameters instead of a fixed quantity~\cite{yin2018byzantine}. Next, a weighted mean mechanism is employed to update the global model parameter, subsequently disseminated back to each BS. Our extensive evaluations, conducted on real-world datasets, demonstrate that the proposed defensive mechanism substantially mitigates the impact of model poisoning attacks on WTP, thereby showcasing a promising avenue for securing FL-based WTP systems against Byzantine attacks.

The contribution of this work is summarized in three folds:
\begin{enumerate}
    \item We present a novel model poisoning attack, employing fake BSs for traffic injection into FL-based WTP systems under a minimum-knowledge scenario.
    \item Conversely, we propose an effective defense strategy designed against various model poisoning attacks, which dynamically trims an adaptive number of model parameters by leveraging the percentile estimation technique.
    \item Lastly, we evaluate both the proposed poisoning attack and the defensive mechanism using real-world traffic datasets from Milan City, where the results demonstrate that the FTI attack indeed compromises FL-based WTP systems, and the proposed defensive strategy proves notably more effective than other baseline approaches.
\end{enumerate}

\section{Related Works and Preliminaries}
\label{sec:related}

\subsection{FL-based WTP}

Consider a wireless traffic forecasting system that employs FL and incorporates a central server located in a macrocell station along with $n$ small-cell BSs (e.g., gNB).
Every BS $i \in [n]$ possesses its own private training dataset $ u_i = \{u^1_i, u^2_i, \ldots, u^M_i\} $, where $ M $ represents the total count of time intervals, and $ u^m_i $ denotes the traffic load on BS $i$ during the $m$-th interval, with $m \in [M]$.
To delineate a prediction model, we construct a series of input-output pairs $\{a_i^j, b_i^j\}_{j=1}^{z}$. Here, each $ a_i^j $ represents a historical subset of traffic data that correlates to its associated output 
$ b_i^j =\{u_i^{m-1}, \dots, u_i^{m-r}, u_i^{m-\omega 1},\dots, u_i^{m-\omega s}\}$.
The parameters $ r $ and $ s $ serve as sliding windows capturing immediate temporal dependencies and cyclical patterns, respectively. Furthermore, $ \omega $ encapsulates inherent periodicities within the network, potentially driven by diurnal user patterns or systematic service demands.
Given the importance of real-time responsiveness in wireless networks, our prediction model is designed for a one-step-ahead forecast. To be specific, for the $i$-th BS, we seek to predict the traffic load $\tilde{b}^j_i$ based on the historical traffic data $a^j_i$ and model parameter $\bm{\theta}$ as
$
\tilde{b}^j_i = f(a^j_i, \bm{\theta}),
$
where $f(\cdot)$ is the regression function.

In a FL-based WTP system, the objective is to minimize prediction errors across $n$ BSs. This can be formulated as the following optimization problem to determine the optimal global model $\bm{\theta}^*$ in the central server:
\begin{align}
\label{all_obj}
\bm{\theta}^* = \arg\min_{\bm{\theta}}  \frac{1}{nz} \sum_{i=1}^{n} \sum_{j=1}^{z} F(f(a_i^j, \bm{\theta}), b_i^j),
\end{align}
where $F$ is the quadratic loss, i.e., $F(f(a_i^j, \bm{\theta}), b_i^j) = \left| f(a_i^j, \bm{\theta}) - b_i^j \right|^2$. Eq.~(\ref{all_obj}) can be resolved in a distributed fashion based on FL with the following three steps in each global training round $t$.

\begin{itemize}
    \item \textbf{Step I (Synchronization).}
    The central server sends the current global model $\bm{\theta}^t$ to all BSs.

    \item \textbf{Step II (Local model training).}
    Each BS $i \in [n]$ utilizes its private time-series training data along with the current global model to refine its own local model, then transmits the updated local model $\bm{\theta}_i^{t}$ back to the server.

    \item \textbf{Step III (Local models aggregation).}
    The central server leverages the aggregation rule ($\text{AR}$) to merge the $n$ received local models and subsequently updates the global model as follows:
    \begin{align}
    \bm{\theta}^{t+1} = \text{AR}\{\bm{\theta}_1^{t}, \bm{\theta}_2^{t}, \ldots, \bm{\theta}_n^{t}\}.
    \end{align}
     The commonly used aggregation rule is the FedAvg~\cite{mcmahan2017communication}, where the server simply averages the received $n$ local models from distributed BSs, i.e., $\text{AR}\{\bm{\theta}_1^{t}, \bm{\theta}_2^{t}, \ldots, \bm{\theta}_n^{t}\} = \frac{1}{n}\sum\limits_{i=1}^n \bm{\theta}_i^{t}$.

\end{itemize}

\subsection{Byzantine-robust Aggregation Rules}
In non-adversarial scenarios, the server aggregates the received local model updates by straightforwardly averaging them~\cite{mcmahan2017communication}. Nevertheless, recent research~\cite{blanchard2017machine} has revealed that this averaging-based aggregation method is susceptible to poisoning attacks, where a single malicious BS can manipulate the final aggregated outcome without constraints. To counteract such potential threats, various Byzantine-robust aggregation rules have been suggested~\cite{blanchard2017machine,yin2018byzantine,cao2020fltrust,fung2018mitigating,sharma2023flair,xia2019faba,fang2022aflguard,xu2024robust}.
For instance, in the Krum method~\cite{blanchard2017machine}, each client's update is scored based on the sum of Euclidean distances to other clients' updates. The global update is then updated by selecting the update from the client (i.e., BS) with the minimum score.
In a Median aggregation scheme~\cite{yin2018byzantine}, the server calculates the median value for each dimension using all the local model updates.
In the FLTrust~\cite{cao2020fltrust}, it is assumed that the server possesses a validation dataset. The server maintains a model derived from this dataset. To determine trust levels, the server computes the cosine similarity between its model update and the update of each BS. These scores are then used to weigh the contribution of each BS to the final aggregated model.

\subsection{Poisoning Attacks to FL-based Systems}

The decentralized nature of FL makes our considered problem susceptible to Byzantine attacks~\cite{fang2020local,cao2022mpaf,shejwalkar2021manipulating,zheng2022poisoning,tolpegin2020data,yin2024poisoning}, where attackers with control over malicious BSs can compromise the FL-based WTP system. Malicious BSs can corrupt their local training traffic data or alter their local models directly.
For instance, in the Trim attack~\cite{fang2020local}, the attacker intentionally manipulates the local models on malicious BSs to cause a significant deviation between the aggregated model after attack and the one before attack.
In the Model Poisoning Attack based on Fake clients (MPAF) attack~\cite{cao2022mpaf}, each malicious BS first multiplies the global model update synchronized from the central server by a negative scaling factor and subsequently transmits these scaled model updates to the server.
In the Random attack~\cite{fang2020local}, every malicious BS randomly generates a vector from a Gaussian distribution and transmits it to the server.
Recently, \cite{zheng2022poisoning} introduced poisoning attacks for FL-based WTP systems,
where the attacker controls some deployed BSs, each with its own local training data. These malicious BSs fine-tune their local models using their respective training data. Subsequently, the attacker scales the local model updates on malicious BSs by applying a scaling factor and sends the scaled model updates to the server. 

However, existing attacks suffer from
the practical implementation limitations. For instance, the attack described in~\cite{zheng2022poisoning} is not feasible because it is based on the unrealistic assumption that an attacker can readily take control of authentic BSs. In reality, it is highly challenging for an attacker to gain such influence over existing, authentic BSs. In the MPAF attack, which has a simpler threat model, the model updates from fake clients are exaggerated by a factor such as $10^6$. This approach is impractical because the central server can easily identify these excessive updates as anomalies and discard them. By contrast, our proposed poisoning attack involves carefully crafting model updates on fake BSs by addressing a parametric optimization problem. This ensures that the server is unable to differentiate these fake updates from benign ones, allowing the attacker to simultaneously breach the integrity of the system without detection.
\begin{figure}[h]
    \centering
    \includegraphics[scale = 0.45]{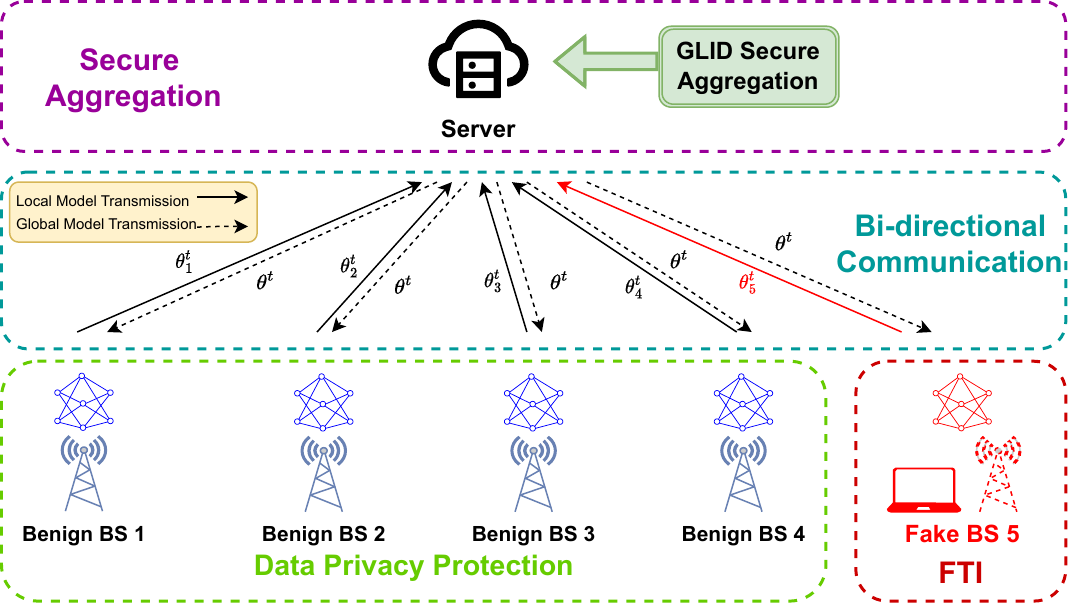}
    \caption{Framework of Security Protection in FL-based WTP.}
    \label{fig:framework}
\end{figure}

\section{Threat Model to FL-based WTP Systems}
\label{sec:threat}

In this section, we present a novel model poisoning attack, employing fake BSs for traffic injection into FL-based WTP systems under a minimum-knowledge scenario.

\subsection{Attacker's Goal}
The attacker's primary goal in compromising the integrity of the FL-based WTP system is to degrade the final global model's performance. This degradation directly impacts the accuracy of real-time traffic predictions, which is a critical aspect of network management and resource allocation. In practical cellular systems, inaccurate traffic predictions can lead to network congestion, poor quality of service, and inefficient use of resources, thereby causing substantial operational challenges for network providers. This disruption not only affects service providers but also has a cascading effect on end-users who rely on consistent and reliable network services.

\vspace{-0.15cm}
\subsection{Attacker's Capability}
The attacker achieves this objective by introducing fake BSs into the targeted FL-based WTP system, as shown in Fig.~\ref{fig:framework}. These fake BSs, which could be simple network devices, mimic the traffic processing behaviors of benign BSs with minimal effort and expense. Unlike the methods proposed in~\cite{zheng2022poisoning} which involve compromising genuine BSs, the use of fake BSs is far more feasible in real-world contexts. Creating fake BSs with open-source projects or emulators~\cite{simulatefake,NoxPlayer,bluestacks,cao2022mpaf} is a low-cost approach that can be executed without the need for sophisticated hacking skills or deep access to the network infrastructure. This approach is particularly viable given the heightened security measures in modern networks, which make compromising genuine BSs increasingly challenging.

\vspace{-0.1cm}
\subsection{Attacker's Knowledge}
The attacker's minimal knowledge about the targeted FL-based WTP system significantly increases the difficulty of executing the attack. In many real-world systems, gaining detailed insights into the central server's aggregation rules or acquiring information about benign BSs is highly challenging due to stringent security protocols and encryption. Therefore, an attack strategy that requires limited knowledge is not only more realistic but also more likely to get undetected. The fake BSs' operation, which is limited to receiving the global information and sending malicious updates, can be executed with basic technical skills, further lowering the barrier to entry for potential attackers. This aspect opens the door to a broader range of network adversaries, including those with limited technical expertise or computing resources.
\vspace{-0.1cm}
\subsection{Fake Traffic Injection Attack}
\label{sec:attack}

The proposed Algorithm~\ref{algo:attack}, referred to as the Fake Traffic Injection (FTI), outlines a Byzantine model poisoning attack strategy designed to manipulate the prediction accuracy of an FL-based WTP system under the aforementioned assumptions. 

Central to the FTI attack is an iterative process where each iteration involves a thorough examination of current global model $\bm{\theta}^{t}$ and base model $\hat{\bm{\theta}}$. For each fake BS $i$, a malicious local model \( \bm{\theta}_i^{t} \) is constructed by combining the global model \( \bm{\theta}^{t} \) and a base model \( \hat{\bm{\theta}} \) in a weighted manner (Line 5).
Following the creation of \( \bm{\theta}_i^{t} \), it evaluates its divergence from the global model using the Euclidean norm (Line 7).
The algorithm then checks for an increase in this distance relative to the prior measurement (Line 8). If the distance has increased, indicating that the malicious local model \( \bm{\theta}_i^{t} \) from some BS is diverging further from the global model \( \bm{\theta}^{t} \) in the central server, the value of \( \eta \) is adjusted upwards. Conversely, if no increase in distance is observed, \( \eta \) is adjusted downwards. The adjustment of \( \eta \) is done in half-steps of its initial value (Lines 8-12). In other words, the value of \( \eta \) indicates the severity of poisoning attacks, measuring their impact or intensity.

\begin{figure}
    \centering
    \includegraphics[scale = 0.28]{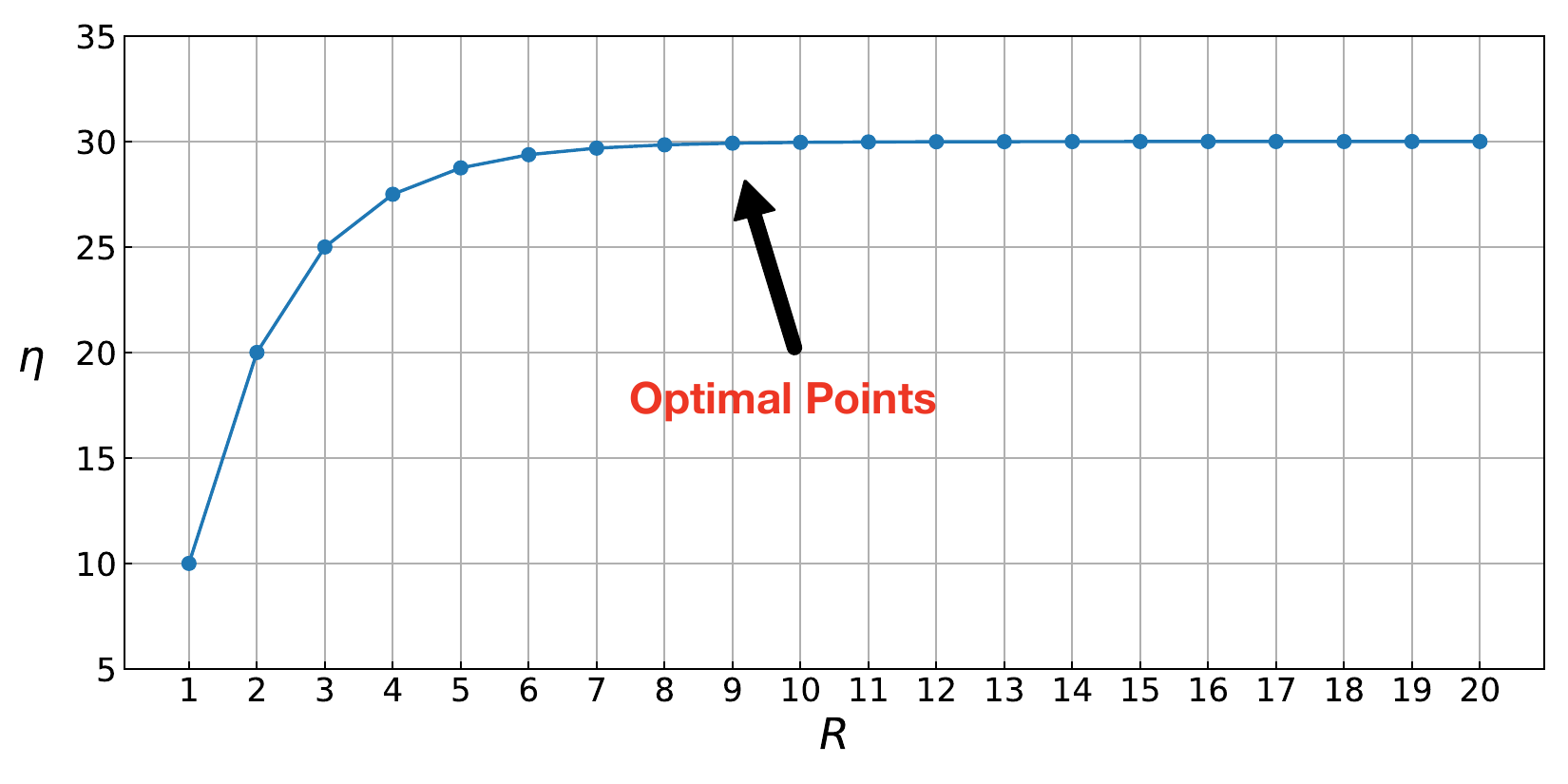}
    \caption{Optimal value of $\eta$ over communication round of $R$ in Algorithm~\ref{algo:attack}.
    }
    \label{fig:eta_r}
\end{figure}

To this end, the algorithm involves guiding the global model to align more closely with a predefined base model in each round. Specifically, during the \( t \)-th round, fake BSs calculate the direction of local model updates, determined by the difference between current global model and base model, denoted as \(  \bm{H}= \hat{\bm{\theta}} - \mathbf{\bm{\theta}}^t \). Moving towards this direction indicates that the global model is becoming more similar to the base model. 
A simple approach to acquire the local model of fake BS involves multiplying $\bm{H}$ by a scaling factor $\eta$. However, this direct method produces sub-optimal attack performance. 
Suppose $n$ is the number of benign BSs, and the attacker wants to inject $m$ fake BSs into the network system.
We propose a method for calculating $\bm{\theta}_i^{t}$ for each fake BS $ i \in [n+1, n+m]$:
\begin{equation}
\bm{\theta}^t_i =   \eta \hat{\bm{\theta}} +  (1 - \eta )\bm{\theta}^{t}.
\end{equation}
\noindent In such cases, an attacker tends to choose a higher value for \( \eta \) to ensure the sustained effectiveness of the attack, as shown in Fig.~\ref{fig:eta_r} with an initial \( \eta \) of 10. This holds true even after the server consolidates the manipulated local updates from fake BSs with legitimate updates from benign BSs.

\begin{algorithm}[t]
\caption{Fake Traffic Injection (FTI)}
\begin{algorithmic}[1]
\Require Current global model $\bm{\theta}^{t}$, base model $\hat{\bm{\theta}}$, \( n \) benign BSs, \( m \) fake BSs, \( \eta \)
\Ensure Fake models \( \bm{\theta}_i^{t}, i \in [n+1, n+m] \)
\State \( \text{step} \gets \eta \)
\State \( \text{PreDist} \gets -1 \)
\For{\( r = 1, 2, \ldots, R \)}
    \For{each fake BS \( i \)}
        \State \( \bm{\theta}_i^{t} \gets  \eta \hat{\bm{\theta}} -  (\eta -1)\bm{\theta}^{t} \)
    \EndFor
    \State \( \text{Dist} \gets \left\| \bm{\theta}_i^{t} - {\bm{\theta}^t}\right\|_2 \)
    \If{\( \text{PreDist} < \text{Dist} \)}
        \State \( \eta \gets \eta+ \frac{\text{step}}{2} \)
    \Else
        \State \( \eta \gets \eta - \frac{\text{step}}{2} \)
    \EndIf
    \State \( \text{step} \gets \frac{\text{step}}{2} \)
    \State \( \text{PreDist} \gets \text{Dist} \)
\EndFor
\State \Return \( \bm{\theta}_i^{t}, i \in [n+1, n+m] \)
\end{algorithmic}
\label{algo:attack}
\end{algorithm}

\section{Global-Local Inconsistency Detection}
\label{sec:defense}

The defense against model poisoning attacks on the FL-based WTP system relies on an aggregation protocol designed to identify malicious BSs. This protocol, named the Global-local Inconsistency Detection (GLID) method, is detailed in Algorithm~\ref{algo:defense}.
In each global round $t$,
GLID primarily scrutinizes the anomalies present in each dimension of the model parameters \( \bm{\bm{\theta}}_i^t \), aiding in the identification of any potentially malicious entities, where $i \in [1, n+m]$, and $n+m$ is the total number of BSs in the system. Such a robust and versatile nature allows the network to adapt to various operational contexts without requiring intricate similarity assessments as in other existing works, like FLTrust~\cite{cao2020fltrust}.

Specifically, GLID approach enhances the detection of potential malicious activities within the network by employing \textit{percentile-based trimming} on each dimension of the model parameters. To establish an effective percentile pair for identifying abnormalities, four statistical methods can be adopted: Standard Deviation (SD), Interquartile Range (IQR), Z-scores, and One-class Support Vector Machine (One-class SVM). 
Suppose the total count of dimensions of model parameter is $D$, then
for the \textbf{default SD method}, the percentile pair for each dimension \( d \) can be calculated as follow:
\begin{equation}
    \text{percentile pair}^t_{d} = \left( g\left( \bar{\bm{\theta}}_{d}^t - k \cdot \sigma^t_d\right), \, g\left( \bar{\bm{\theta}}_{d}^t + k \cdot \sigma^t_d \right) \right),
\end{equation}
where \( \bar{\bm{\theta}}_{d}^t \) is the mean of the \( d \)-th dimension across all models in the $t$-th global training round, \( \sigma^t_d \) is the standard deviation of the $d$-th dimension, and \( k \) is a predefined constant dictating the sensitivity of outlier detection. $g(\cdot)$ is the interpolation function based on standard deviation bound to estimate percentile pairs, shown as follows:
\begin{equation}
    g(x) = \left( \frac{P(x) - 0.5}{n+m} \right) \times 100,
\end{equation}
where $P(x)$ is the position of $x$ in the sorted dataset.
We use \( k=3 \) for general purposes. Given that different tasks may require varied percentile bounds, a precise estimation method is crucial for generalizing our defense strategy. The detailed percentile estimation methods can be found later in this section. In the FL-based WTP system, model parameters in the \( d \)-th dimension exceeding these percentile limits are flagged as malicious, and their weights \( \alpha^t_i \) are assigned as 0. The other benign values in this dimension are aggregated using a weighted average rule, where 
the weights \( \alpha^t_{d,i} \) are inversely proportional to the absolute deviation of each value \( \bm{\theta}^t_{d,i} \) from the mean \( \bar{\bm{\theta}}_{d}^t \), and normalized by the standard deviation \( \sigma^t_d \).
It can be represented as follows:
\begin{equation}
    \alpha^t_{d,i} = \frac{\sigma^t_d}{\left| \bm{\theta}^t_{d,i} - \bar{\bm{\theta}}_{d}^t \right|}.
\end{equation}
These weights of the \( d \)-th dimension are then normalized and applied to aggregate each BS's local model \( \bm{\theta}_i^t \) into a global model \( \bm{\theta}^{t+1} \), which can be represented as follow in the view of each dimension:
\begin{equation}
    \bm{\theta}^{t+1}_d = \frac{\sum_{i=1}^{n+m} \alpha^t_{d,i} \cdot \bm{\theta}^t_{d,i}}{\sum_{i=1}^{n+m} \alpha^t_{d,i}}.
\end{equation}
Subsequently, the server broadcasts this aggregated global model parameter $\bm{\theta}^{t+1}$ back to all BSs for synchronization.

There are three additional percentile estimation strategies listed below. Based on the upper and lower bound computed below, we can get a final percentile estimation decision to detect abnormal values in each dimension.

\begin{itemize}
    \item \textbf{Interquartile Range (IQR)}: 
    The IQR method calculates the range between the first and third quartiles (25$^{\rm th}$ and 75$^{\rm th}$ percentiles) of the data, identifying outliers based on this range. For each dimension \( d \), the outlier bounds are:
    \begin{align}
        \text{lower bound}^t_{d,\text{IQR}} &= Q1^t_d - k_{\text{IQR}} \cdot \text{IQR}^t_d, \\
        \text{upper bound}^t_{d,\text{IQR}} &= Q3^t_d + k_{\text{IQR}} \cdot \text{IQR}^t_d,
    \end{align}
    where \( Q1^t_d \) and \( Q3^t_d \) are the first and third quartiles, and \( k_{\text{IQR}} \) adjusts sensitivity.

    \item \textbf{Z-scores}: 
    The Z-score method measures how many standard deviations a point is from the mean. For each dimension \( d \), the normal range bounds are:
    \begin{align}
        \text{lower bound}^t_{d,\text{Z-score}} &= g\left( \bar{\bm{\beta}}_{d}^t - k_{\text{Z}} \cdot \sigma^t_d\right), \\
        \text{upper bound}^t_{d,\text{Z-score}} &= g\left( \bar{\bm{\beta}}_{d}^t + k_{\text{Z}} \cdot \sigma^t_d \right),
    \end{align}
    where \( k_{\text{Z}} \) is the number of standard deviations for the normal range.

    \item \textbf{One-Class SVM}: 
    One-Class SVM constructs a decision boundary for anomaly detection. The decision function for each dimension \( d \) is:
    \begin{align}
        f^t_d(\bm{\beta}) &= \text{sign}\left( \sum_{i=1}^{n_{\text{SV}}} \gamma_i \cdot K(\bm{\beta}^t_{\text{SV}_i, d}, \bm{\beta}) - \rho \right), 
    \end{align}
    where $\bm{\beta}^t_{\text{SV}_i, d}$ are the support vectors, $\gamma_i$ are the Lagrange multipliers, $K(\cdot, \cdot)$ is the kernel function, and $\rho$ is the offset. A point $\bm{\beta}$ is an outlier if  $f^t_d(\bm{\beta}) < 0$. 

\end{itemize}

In essence, this defense mechanism is a strategic amalgamation of direct statistical trimming and aggregation, targeting the preservation of the global model's integrity against poisoning attacks. By accurately isolating and excluding malicious BSs prior to model aggregation process, it significantly diminishes the likelihood of adversarial disruption in the FL framework. Additionally, its capacity to accommodate various dimensions and adapt to different inconsistency metrics and aggregation protocols considerably extends its applicability across a broad spectrum of distributed wireless network scenarios.

\begin{algorithm}[t]
\caption{Global-local Inconsistency Detection (GLID)}
\begin{algorithmic}[1]
\Require Local models \( \bm{\theta}_1^t, \bm{\theta}_2^t, \ldots, \bm{\theta}_{n+m}^t \), current global model \( \bm{\theta}^t \), \(k\)
\Ensure Aggregated global model \( \bm{\theta}^{t+1} \)

\For{\( d = 1, 2, \ldots, D \)}
    \State \( \bar{\bm{\theta}}_{d}^t \gets \frac{1}{n+m} \sum_{i=1}^{n+m} \bm{\theta}^t_{d,i} \)  
    \State \( \sigma^t_d \gets \sqrt{\frac{1}{n+m} \sum_{i=1}^{n+m} (\bm{\theta}^t_{d,i} - \bar{\bm{\theta}}_{d}^t)^2} \)  
    \State \( \text{percentile}^t_d \gets \left( g\left( \bar{\bm{\theta}}_{d}^t - k \cdot \sigma^t_d\right), \, g\left( \bar{\bm{\theta}}_{d}^t + k \cdot \sigma^t_d \right) \right) \)
    \State Identify malicious BSs based on percentile pairs
    \For{each BS $i$}
        \If{\( \bm{\theta}^t_{d,i} \) is benign}
            \State \( \alpha^t_{d,i} \gets \frac{\sigma^t_d}{\left| \bm{\theta}^t_{d,i} - \bar{\bm{\theta}}_{d}^t \right|} \)
        \Else
            \State \( \alpha^t_{d,i} \gets 0 \)
        \EndIf
    \EndFor
    \State \( \bm{\theta}_d^{t+1} \gets \frac{\sum_{i=1}^{n+m} \alpha^t_{d,i} \cdot \bm{\theta}^t_{d,i}}{\sum_{i=1}^{n+m} \alpha^t_{d,i}} \)
\EndFor
\State \(\bm{\theta}^{t+1} \gets \left[ \bm{\theta}_1^{t+1}, \bm{\theta}_2^{t+1}, \ldots, \bm{\theta}_D^{t+1} \right]\)
\State \Return \( \bm{\theta}^{t+1} \)
\end{algorithmic}
\label{algo:defense}
\end{algorithm}

\section{Evaluations}
\label{sec:exp}

In this section, we demonstrate the effectiveness of our FTI poisoning attack and the GLID defense mechanism. Extensive evaluation results are provided regarding the performance metrics in multiple dimensions. 

\vspace{-0.1cm}
\subsection{Experiment Setup}

\subsubsection{Datasets}
We utilize the real-world datasets obtained from Telecom Italia~\cite{dataset} to evaluate our proposed methods. The wireless traffic data in Milan is segmented into 10,000 grid cells, with each cell served by a BS covering an area of approximately 235 meters on each side. Milan Dataset contains three subset datasets, ``Milan-Internet'', ``Milan-SMS'' and ``Milan-Calls''. These datasets capture different types of wireless usage patterns, and we are mainly focusing on ``Milan-Internet''. Such comprehensive data collection enables an in-depth analysis of urban telecommunication behavior.

\subsubsection{Baseline Schemes}
We evaluate various state-of-the-art model poisoning attacks as comparison points to our proposed FTI attack. Furthermore, we employ these baseline poisoning attacks to highlight the effectiveness of our defense strategy GLID.

\begin{itemize}
    \item \textbf{Trim attack~\cite{fang2020local}:} It processes each key within a model dictionary, computing and utilizing the extremes in a designated dimension to determine a \textit{directed} dimension, where model parameters are selectively zeroed or retained to influence the model behavior.
    
    \item \textbf{History attack~\cite{cao2022mpaf}:} It iterates over model parameters, replacing current values with historically scaled ones, effectively warping the model parameters using past data to misguide the aggregation process.
    
    \item \textbf{Random attack~\cite{cao2022mpaf}:} It disrupts the model by replacing parameters with random and normally distributed values, scaled to maintain a semblance of legitimacy, thereby injecting controlled chaos into the aggregation process.
    
    \item \textbf{MPAF~\cite{cao2022mpaf}:} It calculates a directional vector derived from the difference between initial and current parameters. This vector is then used to adjust model values, intentionally diverging from the model's original trajectory to introduce an adversarial bias. Following these calculations, the fake BSs are injected into the system.
    
    \item \textbf{Zheng attack~\cite{zheng2022poisoning}:} It inverts the direction of model updates by incorporating the negative of previous global updates. This inversion is refined through error maximization, generating a poison that proves challenging to detect due to its alignment with the model's error landscape.
\end{itemize}

Besides, we consider several baseline defensive mechanisms to demonstrate the effectiveness of our attack and defense. 
\begin{itemize}
    \item \textbf{Mean~\cite{mcmahan2017communication}:} It calculates the arithmetic mean of updates in each dimension, assuming equal trustworthiness among all BSs. However, this method is susceptible to the influence of extreme values.
    
    \item \textbf{Median~\cite{yin2018byzantine}:} It identifies the median value in each dimension for each parameter across updates, which inherently discards extreme contributions to enhance the robustness against outliers.
    
    \item \textbf{Trim~\cite{yin2018byzantine}:} It discards a specified percentage of the highest and lowest updates before computing the mean in each dimension, thereby reducing the potential sway of anomalous or malicious updates on the aggregate model.
    
    \item \textbf{Krum~\cite{blanchard2017machine}:} Each BS's update is scored based on the sum of Euclidean distances to other BSs' updates. The global update is then updated by selecting the update from the BS with the minimum score.

    \item \textbf{FoolsGold~\cite{fung2018mitigating}:} It calculates a cosine similarity matrix among all BSs and adjusts the weights for each BS based on these similarities. The weighted gradients are then aggregated to form a global model.

    \item \textbf{FABA~\cite{xia2019faba}:} It computes the Euclidean distance for each BS's model from the mean of all received models. By identifying and excluding a specific percentage of the most distant models, this process effectively filters out potential outliers or malicious updates.

    \item \textbf{FLTrust~\cite{cao2020fltrust}:} Cosine similarity is calculated between the server's current model and each BS's model to generate trust scores. These scores are then used to weigh the BS's contribution to the final aggregated model.

    \item \textbf{FLAIR~\cite{sharma2023flair}:} Each BS calculates ``flip-scores'' derived from the changes in gradient directions and ``suspicion-scores'' based on historical behavior. These scores are used to adjust the weights assigned to each BS's contributions to the global model.
\end{itemize}

\subsubsection{Experimental Settings and Performance Metrics}

In our experimental setup, we randomly selected 100 BSs to evaluate the impact of poisoning attacks and the effectiveness of defense mechanisms. By default, we report the results on Milan-Internet dataset. Model training is configured with a learning rate of 0.001 and a batch size of 64. We inject a 20\% percentage of fake BSs to mimic benign ones in the system for FTI attack and simulate a scenario where 20\% of the BSs are compromised for other baseline attacks.  Our proposed FTI attack utilizes a parameter \( \eta = 10 \), and other attacks utilize a scaling factor of 1000. For the Trim aggregation rule, we discard 20\% of the model parameters from all BSs. In our proposed GLID defense, we employ the standard deviation (SD) method as the default percentile estimation method. Throughout the measurement campaign, we adopt Mean Absolute Error (MAE) and Mean Squared Error (MSE) as the primary metrics for performance evaluation. 
MSE quantifies the average of the squared discrepancies between estimated and actual values, while MAE calculates the average absolute differences across predictions, disregarding their directional errors.
The larger the MAE and MSE, the better the effectiveness of the attack.

\vspace{-0.1cm}
\subsection{Numerical Results}

\subsubsection{Performance of Proposed Methods}

\begin{table*}
\centering
\caption{Performance Metrics for Milan-Internet Dataset}
\label{tab:net-milan}
\begin{tabular}{|c|c|C{1.5cm}|C{1.5cm}|C{1.5cm}|C{1.5cm}|C{1.5cm}|C{1.5cm}|C{1.5cm}|}
\hline
\multirow{2}{*}{Aggregation Rule} & \multirow{2}{*}{Metric} & \multicolumn{7}{c|}{Attack} \\ \cline{3-9}
                                  &                        & NO & Trim & History & Random & MPAF & Zheng & \textit{\textbf{FTI}} \\ \hline
\multirow{2}{*}{Mean}             
                                  & MAE                    & 0.211  & 100.0    & 100.0       & 100.0      & 100.0    & 0.698       & 100.0 \\
                                  & MSE                  & 0.086  & 100.0    & 100.0       & 100.0      & 100.0    & 0.294       & 100.0 \\ \hline
\multirow{2}{*}{Median}           
                                  & MAE                    & 0.211  & 0.213    & 0.211       & 0.212      & 0.211    & 0.217       & 100.0 \\
                                  & MSE                  & 0.086  & 0.086    & 0.087       & 0.086      & 0.086    & 0.095       & 100.0 \\ \hline
\multirow{2}{*}{Trim}             
                                  & MAE                    & 0.211  & 0.212    & 0.212       & 0.211      & 0.212    & 0.239       & 100.0 \\
                                  & MSE                  & 0.086  & 0.087    & 0.089       & 0.086      & 0.088    & 0.106       & 100.0 \\ \hline
\multirow{2}{*}{Krum}            
                                  & MAE                    & 0.221  & 0.225    & 100.0       & 0.225      & 100.0    & 0.225       & 100.0 \\
                                  & MSE                  & 0.091  & 0.093    & 100.0       & 0.094      & 100.0    & 0.094       & 100.0 \\ \hline
\multirow{2}{*}{FoolsGold}        
                                  & MAE                    & 0.213  & 100.0    & 100.0       & 100.0      & 100.0    & 0.934       & 100.0 \\
                                  & MSE                  & 0.095  & 100.0    & 100.0       & 100.0      & 100.0    & 0.607       & 100.0 \\ \hline
\multirow{2}{*}{FABA}             
                                  & MAE                    & 0.219  & 100.0    & 100.0       & 100.0      & 100.0    & 0.623       & 100.0 \\
                                  & MSE                  & 0.089  & 100.0    & 100.0       & 100.0      & 100.0    & 0.249       & 100.0 \\ \hline
\multirow{2}{*}{FLTrust}          
                                  & MAE                    & 0.242  & 0.234    & 100.0       & 0.240      & 100.0    & 3.182       & 100.0 \\
                                  & MSE                  & 0.094  & 0.092    & 100.0       & 0.094      & 100.0    & 1.208       & 100.0 \\ \hline
\multirow{2}{*}{FLAIR}            
                                  & MAE                    & 0.216  & 0.228    & 100.0       & 100.0      & 100.0    & 0.250       & 100.0 \\
                                  & MSE                  & 0.094  & 0.088    & 100.0       & 100.0      & 100.0    & 0.096       & 100.0 \\ \hline
\multirow{2}{*}{\textbf{GLID}}      
                                  & MAE                    & \textbf{0.211}  & \textbf{0.211}    & \textbf{0.212}       & \textbf{0.211}      & \textbf{0.211}    & \textbf{0.212}       & \textbf{72.383} \\
                                  & MSE                  & \textbf{0.086}  & \textbf{0.087}    & \textbf{0.086}       & \textbf{0.086}      & \textbf{0.087}    & \textbf{0.086}       & \textbf{27.528} \\ \hline

\end{tabular}
\vspace{-0.5cm}
\end{table*}

The FTI Attack, in particular, exposes significant vulnerabilities in numerous aggregation methods. It is observed that under our FTI Attack, both Mean and Krum Rules are completely compromised, as reflected by their MAE and MSE values reaching over 100.0 (values exceeding 100 are capped at 100). This result denotes a total breakdown in their WTP functionality. The Median Rule further emphasizes the severity of FTI Attack, with both its MAE and MSE escalating from modest baseline figures to 100. This sharp contrast highlights FTI attack's reliable performance against other defenses, 
such as Trim Attack against Median rule, where the increase in MAE and MSE is relatively minor at 0.234 and 0.092,  respectively. Additionally, 
the Trim Rule, typically considered robust, exhibits a drastic increase in MAE to over 100.0, a significant rise from its baseline without any attack (termed as NO in Table \ref{tab:net-milan}) of 0.211. This surge underscores Trim Rule's vulnerability to the FTI Attack, marking a notable departure from its typical resilience. Similar results can also be found in other aggregation rules under FTI attack, such as FoolsGold, FABA, FLTrust, and FLAIR. The FTI attack has the best overall performance against the given defenses.
The Zheng attack, however, presents a distinct pattern of disruption. When subjected to this attack, FLTrust, which typically exhibits lower error metrics, shows a significant compromise, evidenced by the dramatic increase in its MAE to 3.182 and MSE to 1.208. Such a tailored nature of Zheng attack appears to target specific vulnerabilities within FLTrust, which are not as apparent in other scenarios, such as Trim Attack, where the rise in MAE and MSE for FLTrust is relatively modest. Regarding the MPAF Attack, most aggregation rules in the table do not show a convincing defense, except for a few like Median, Trim, and GLID.

\vspace{-0.06cm}

Next, if we turn our attention to the defender's standpoint, the proposed GLID aggregation method demonstrates consistent performance stability across various attacks. Both its MAE and MSE values remain close to their baseline levels. Even in the case of our FTI attack, GLID manages to keep errors below 100, which is 72.383 and 27.528 for MAE and MSE respectively. This stability is particularly noteworthy, especially when compared to other rules such as FLAIR, which exhibit a significant deviation from their non-attacked baselines under the same adversarial conditions. GLID's ability to sustain its performance in the face of diverse and severe attacks underscores its potential as a resilient aggregation methodology.

\subsubsection{Evaluation on the Impact of $\eta$}

\begin{figure}[ht]
    \centering
    \includegraphics[scale = 0.3]{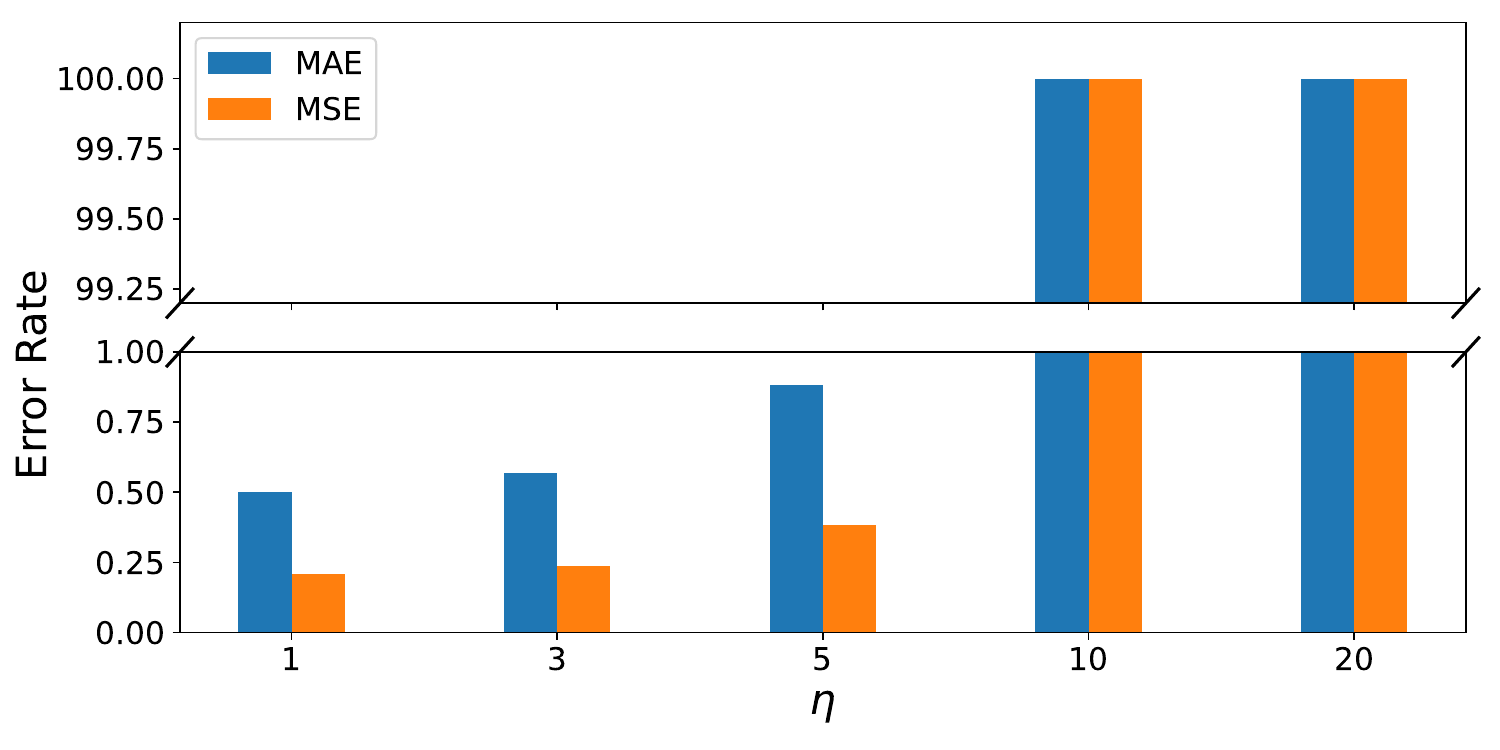}
    \vspace{-0.05cm}
    \caption{Impact of Values of \( \eta \).}
    \label{fig:eta}
    \vspace{-0.25cm}
\end{figure}

The step size \( \eta \) in our proposed FTI attack (see Algorithm~\ref{algo:attack}) serves as a dynamic scaling factor, and its initial value significantly influences the model's performance metrics. This impact is illustrated in Fig.~\ref{fig:eta}, where the Median aggregation rule is employed as the baseline defense strategy. A notable observation is the correlation between increasing values of \( \eta \) and the corresponding rise in MAE and MSE. For example, at \( \eta = 1 \), the MAE and MSE are relatively low, recorded at 0.501 and 0.208, respectively. However, increasing \( \eta \) to higher values, such as 10 or 20, results in a dramatic surge that reaches the maximum error rate. This increase suggests a significant compromise in the model, surpassing the predefined threshold for effective detection of the attack.
The rationale behind this analysis emphasizes the pivotal role of \( \eta \) in determining the \textit{strength} of a poisoning attack. An increased initial \( \eta \) tends to degrade model performance, deviating significantly from its expected operational state. Simultaneously, a higher \( \eta \) also raises the risk of the attack's perturbations being detected and eliminated during the defense process. 

\subsubsection{Evaluation on Percentage of Fake BSs}

The degree of compromise in BSs significantly influences the model's performance, as evidenced in Table~\ref{tab:percentage}. 
By adopting Median aggregation as the defensive approach, the model first exhibits resilience at lower compromise levels, such as with only 5\%--10\% fake BSs in the scenario. 
However, a noticeable decline in performance is observed as the percentage of fake BSs increases to 20\% or higher. This deterioration is evident as MAE and MSE values reach 100.0 in all categories, signaling a complete model failure.
The underlying principle behind this trend suggests the model's limited tolerance to malicious interference. 
More precisely, the network system can withstand below 20\% compromise without significant performance degradation. 
However, beyond this threshold, the model's integrity is severely undermined, resulting in a complete system breakdown. This observation highlights the critical importance of implementing robust security measures to prevent excessive compromise of BSs, ensuring the model's reliability and effectiveness.

\begin{table}[ht]
    \centering
    \caption{Impact of Percentages of Fake BSs}
    \label{tab:percentage}
    \scriptsize
    \begin{tabular}{|c|c|c|c|c|c|c|c|}
    \hline
    \multirow{2}{*}{Pct.} & \multirow{2}{*}{Metric} & \multicolumn{6}{c|}{Attack} \\
    \cline{3-8}
                          &                         & {Trim} & {Hist.} & {Rand.} & {MPAF} & {Zhe.} & {\textbf{FTI}} \\
    \hline
    \multirow{2}{*}{5\%}  & MAE                     & 0.221 & 0.215 & 0.219 & 0.215 & 0.213 & \textbf{0.229} \\
                          & MSE                     & 0.088 & 0.089 & 0.088 & 0.088 & 0.088 & \textbf{0.089} \\
    \hline
    \multirow{2}{*}{10\%} & MAE                     & 0.220 & 0.213 & 0.218 & 0.213 & 0.214 & \textbf{0.258} \\
                          & MSE                     & 0.087 & 0.090 & 0.088 & 0.090 & 0.096 & \textbf{0.104} \\
    \hline
    \multirow{2}{*}{20\%} & MAE                     & 0.223 & 0.218 & 0.218 & 0.216 & 0.269 & \textbf{100.0} \\
                          & MSE                     & 0.087 & 0.096 & 0.087 & 0.092 & 0.136 & \textbf{100.0} \\
    \hline
    \multirow{2}{*}{30\%} & MAE                     & 100.0 & 100.0 & 100.0 & 100.0 & 5.990 & \textbf{100.0} \\
                          & MSE                     & 100.0 & 100.0 & 6.141 & 100.0 & 1.154 & \textbf{100.0} \\
    \hline
    \multirow{2}{*}{40\%} & MAE                     & 100.0 & 100.0 & 100.0 & 100.0 & 100.0 & \textbf{100.0} \\
                          & MSE                     & 100.0 & 100.0 & 100.0 & 100.0 & 100.0 & \textbf{100.0} \\
    \hline
    \end{tabular}
\end{table}
\subsubsection{Evaluations on Percentile Estimation Methods}

The dynamic trimming of an adaptive number of model parameters through percentile estimation, which is adapted in GLID, is effective for an effective defense strategy against various model poisoning attacks. In the comparative analysis of various estimation methods, as shown in Table~\ref{tab:estimation-method}, Standard Deviation (SD) estimation emerges as the best technique, exhibiting a marked consistency and robustness across a spectrum of estimation approaches. This is evidenced by the consistently low MAE and MSE values for SD across these approaches, at 0.219 and 0.087, respectively. 
In contrast, other methods have varying degrees of inconsistency and vulnerability. 
For instance, One-class SVM exhibits pronounced variability, with MAE and MSE values reaching the maximal error level of over 100.0 under Trim, History, and MPAF attacks. 
Such a disparity in performance, particularly the stably lower error rates of SD compared to the significant fluctuations in other estimation methods, positions SD as a reliable and effective percentile estimation technique in GLID. 

\begin{table}
\centering
\caption{Impact of Percentile Estimation Methods}
\label{tab:estimation-method}
\setlength{\tabcolsep}{4pt}
\scriptsize
\begin{tabular}{|c|c|c|c|c|c|c|c|c|}
\hline
\multirow{2}{*}{Method}   & \multirow{2}{*}{Metric} & \multicolumn{7}{c|}{Attack} \\ \cline{3-9}
                          &                         & NO & Trim & Hist. & Rand. & MPAF & Zhe. & \textbf{FTI} \\
\hline
\multirow{2}{*}{SD}       & MAE                     & 0.219 & 0.219 & 0.219 & 0.218 & 0.219 & 0.219 & \textbf{72.38} \\
                          & MSE                     & 0.087 & 0.087 & 0.087 & 0.087 & 0.087 & 0.087 & \textbf{27.52} \\ 
\hline
\multirow{2}{*}{IQR}      & MAE                     & 0.219 & 0.220 & 0.220 & 0.219 & 0.210 & 0.218 & \textbf{100.0} \\
                          & MSE                     & 0.087 & 0.087 & 0.087 & 0.087 & 0.087 & 0.088 & \textbf{100.0} \\
\hline
\multirow{2}{*}{Z-scores} & MAE                     & 0.219 & 0.219 & 0.219 & 0.219 & 0.220 & 1.047 & \textbf{100.0} \\
                          & MSE                     & 0.087 & 0.087 & 0.088 & 0.087 & 0.087 & 0.401 & \textbf{100.0} \\
\hline
\multirow{2}{*}{SVM}      & MAE                     & 0.219 & 100.0 & 100.0 & 0.220 & 100.0 & 0.713 & \textbf{100.0} \\
                          & MSE                     & 0.087 & 100.0 & 100.0 & 0.087 & 100.0 & 0.275 & \textbf{100.0} \\
\hline
\end{tabular}
\end{table}

\vspace{-0.05cm}
\subsubsection{Evaluations on the Impact of BS Density}

Given the percentage of fake BSs at 20\%, 
Figs. 4(a)-(d) compare Median and GLID rules with varying densities of BS in the network scenario. It is interesting to see that the total number of BSs does not significantly impact the performance of any attack and defense mechanisms, especially for our FTI and GLID. Under Median aggregation, FTI consistently shows maximal error (MAE and MSE at over 100.0) across different BS densities, indicating a failure of the defense. 
This consistent pattern of stable performance across varying participants in the FL-based WTP system suggests that the total number of BS does not substantially influence the effectiveness of the attack and defense strategies.

\begin{figure*}
    \centering
    \begin{subfigure}[b]{0.37\textwidth}
        \centering
        \includegraphics[width=\linewidth]{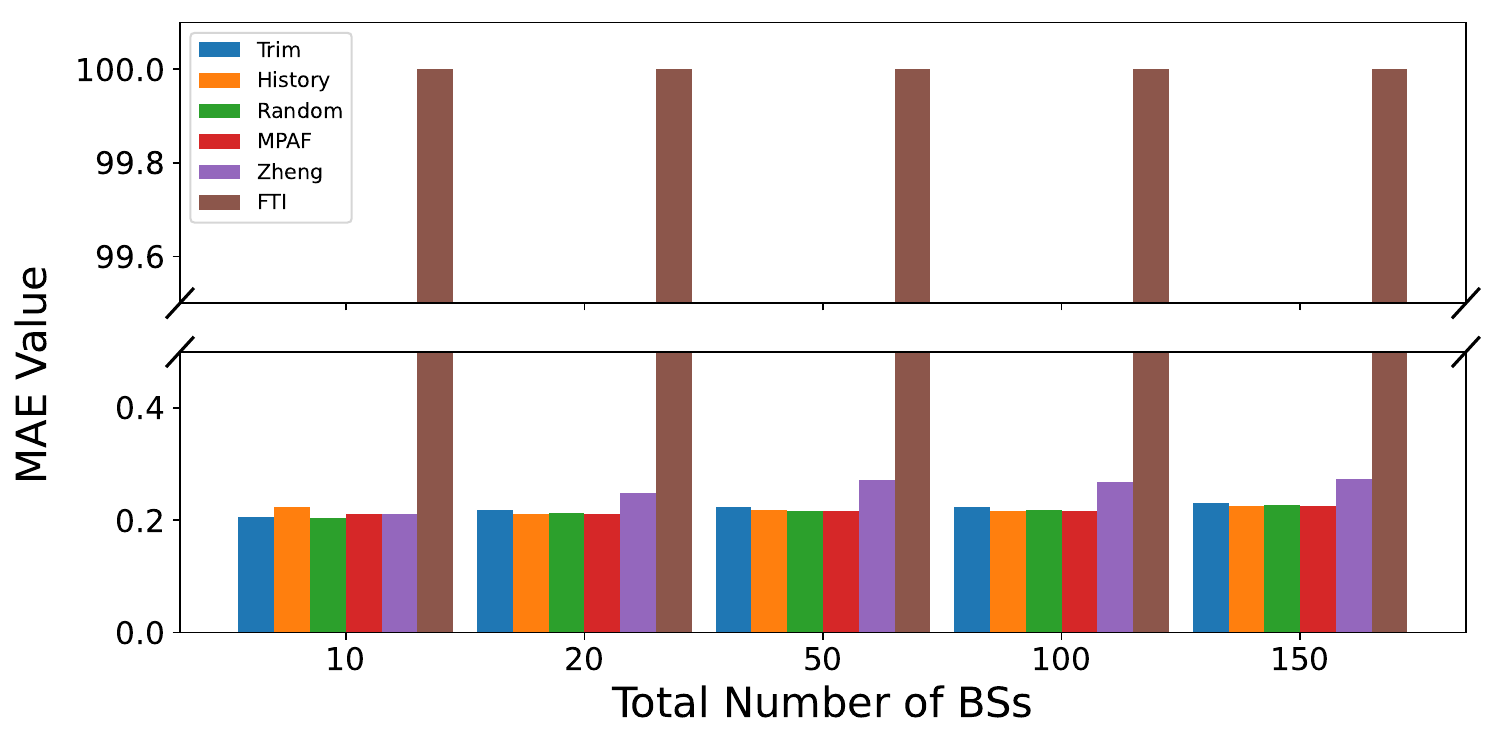}
        \caption{Median AR w.r.t MAE}
        \label{fig:num_mae}
    \end{subfigure}
    \hspace{1pt}
    \begin{subfigure}[b]{0.37\textwidth}
        \centering
        \includegraphics[width=\linewidth]{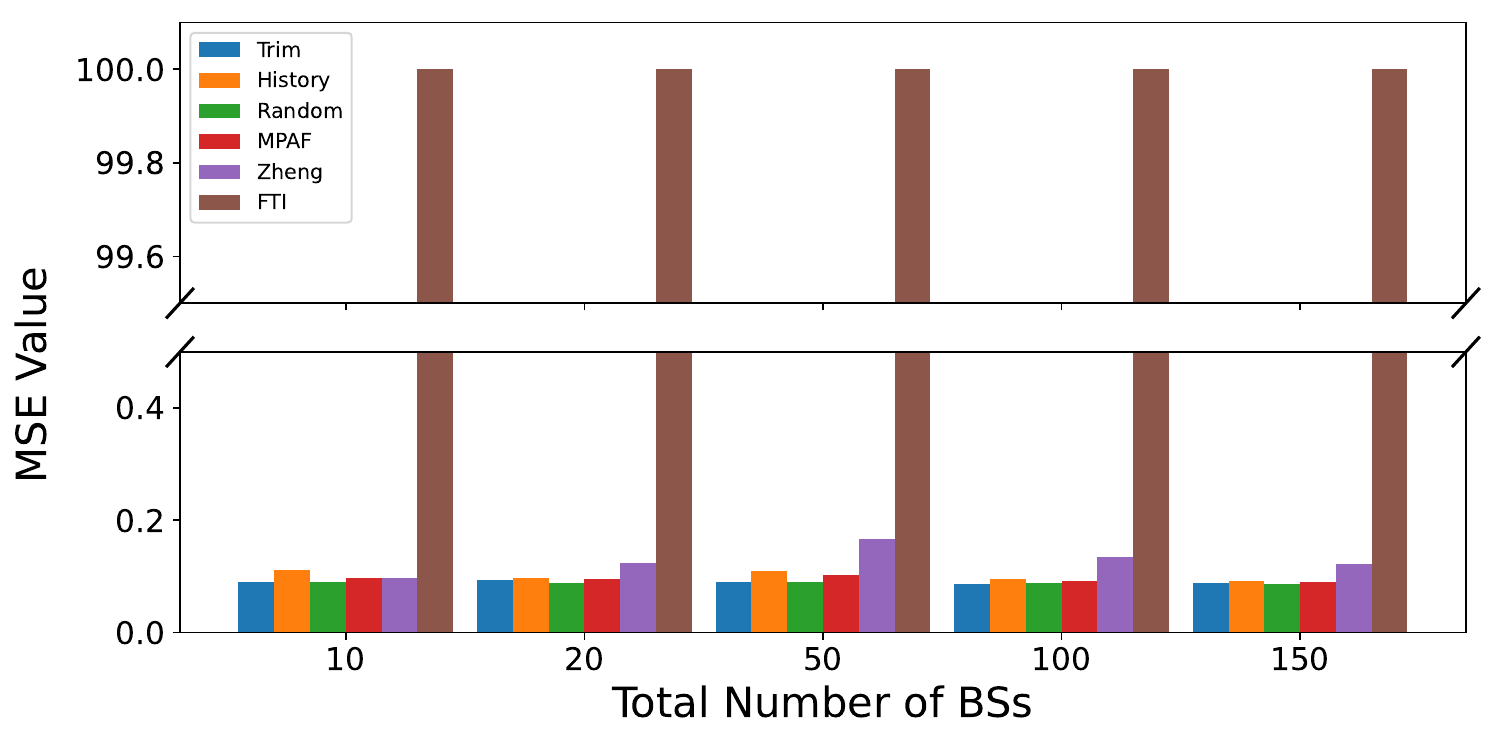}
        \caption{Median AR w.r.t MSE}
        \label{fig:num_mse}
    \end{subfigure}
    \begin{subfigure}[b]{0.37\textwidth}
        \centering
        \includegraphics[width=\linewidth]{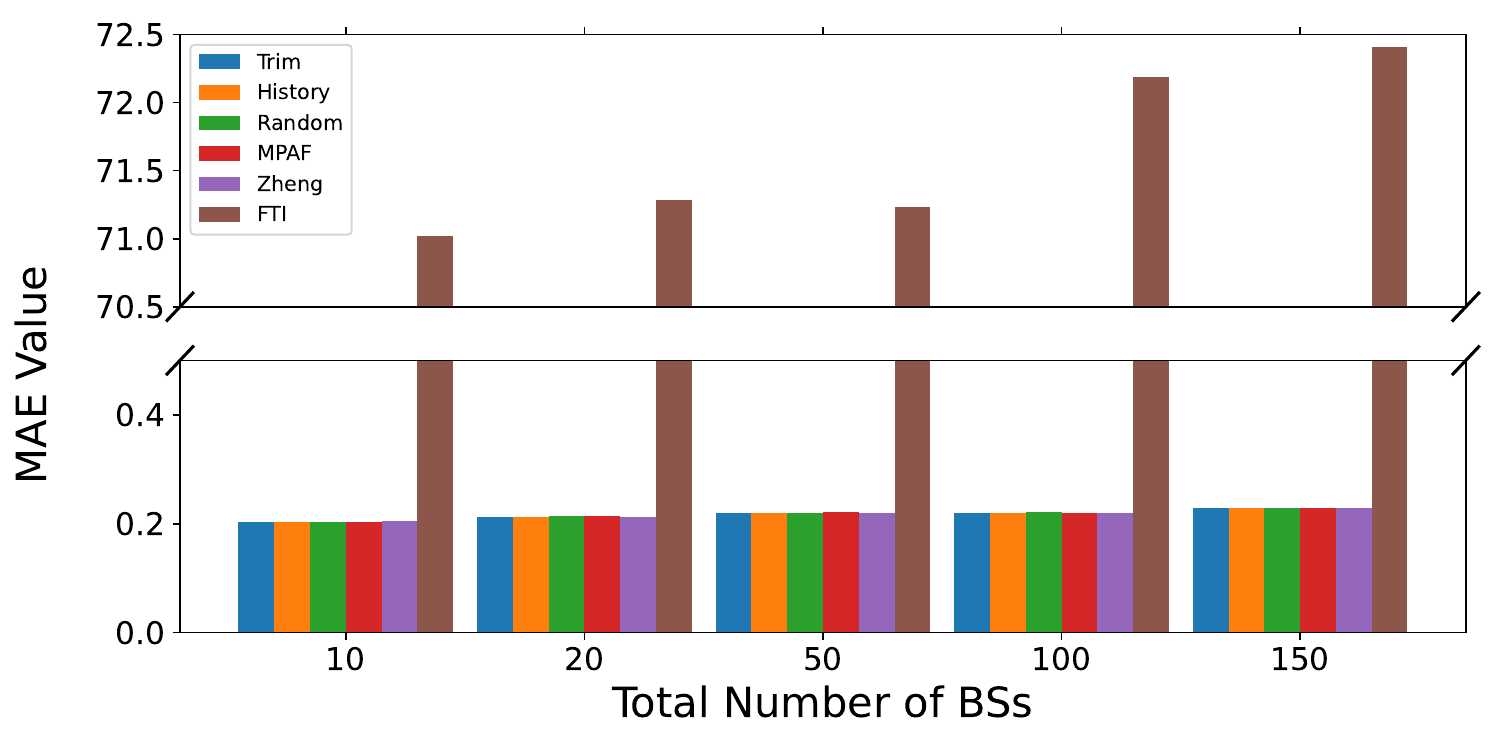}
        \caption{GLID AR w.r.t MAE}
        \label{fig:glid_mae}
    \end{subfigure}
    \hspace{1pt}
    \begin{subfigure}[b]{0.37\textwidth}
        \centering
        \includegraphics[width=\linewidth]{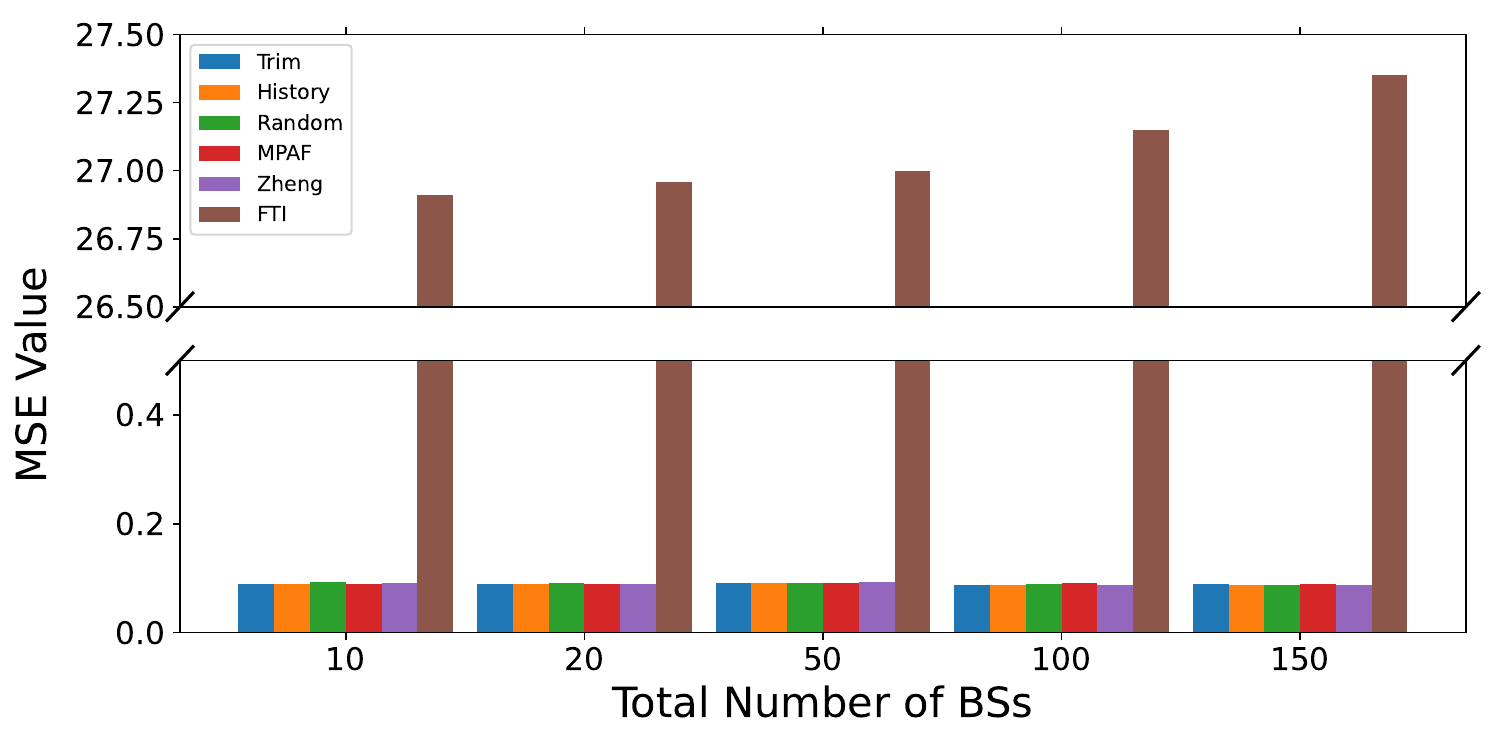}
        \caption{GLID AR w.r.t MSE}
        \label{fig:glid_mse}
    \end{subfigure}

    \caption{The impact of BS density on the performance of Median and GLID methods with respect to MAE and MSEs.}
    \label{fig:combined_figures}
        \vspace{-0.5cm}
\end{figure*}

\subsubsection{Evaluations on the Percentile Range of GLID}
\label{sec:percentile}

\begin{table}[ht]
\centering
\vspace{-0.1cm}
\caption{Impact of Different Percentile Pairs}
\label{tab:percentile}
\scriptsize
\begin{tabular}{|c|c|c|c|c|c|c|c|}
\hline
\multirow{2}{*}{Pair} & \multirow{2}{*}{Metric} & \multicolumn{6}{c|}{Method} \\ \cline{3-8}
                      &                         & Trim & Hist. & Rand. & MPAF & Zhe. & \textbf{FTI} \\ \hline
\multirow{2}{*}{[10, 70]} & MAE                  & 100.0 & 100.0 & 100.0 & 100.0 & 0.710 & \textbf{100.0} \\
& MSE                  & 100.0 & 100.0 & 100.0 & 100.0 & 0.279 & \textbf{100.0} \\
\hline
\multirow{2}{*}{[20, 70]} & MAE                  & 0.215 & 0.214 & 0.218 & 0.217 & 0.216 & \textbf{100.0} \\
& MSE                  & 0.083 & 0.085 & 0.084 & 0.082 & 0.086 & \textbf{100.0} \\
\hline
\multirow{2}{*}{[30, 70]} & MAE                  & 0.218 & 0.219 & 0.220 & 0.215 & 0.217 & \textbf{72.382} \\
& MSE                  & 0.090 & 0.088 & 0.089 & 0.086 & 0.088 & \textbf{27.246} \\
\hline
\multirow{2}{*}{[10, 80]} & MAE                  & 100.0 & 100.0 & 100.0 & 100.0 & 0.711 & \textbf{100.0} \\
& MSE                  & 100.0 & 100.0 & 100.0 & 100.0 & 0.275 & \textbf{100.0} \\
\hline
\multirow{2}{*}{[20, 80]} & MAE                  & 0.217 & 0.215 & 0.218 & 0.214 & 0.216 & \textbf{72.168} \\
& MSE                  & 0.085 & 0.083 & 0.084 & 0.082 & 0.086 & \textbf{27.147} \\
\hline
\multirow{2}{*}{[30, 80]} & MAE                  & 0.220 & 0.218 & 0.219 & 0.216 & 0.217 & \textbf{71.298} \\
& MSE                  & 0.088 & 0.089 & 0.086 & 0.088 & 0.090 & \textbf{27.022} \\
\hline
\multirow{2}{*}{[10, 90]} & MAE                  & 100.0 & 100.0 & 100.0 & 100.0 & 0.712 & \textbf{100.0} \\
& MSE                  & 100.0 & 100.0 & 100.0 & 100.0 & 0.274 & \textbf{100.0} \\
\hline
\multirow{2}{*}{[20, 90]} & MAE                  & 0.215 & 0.217 & 0.218 & 0.216 & 0.214 & \textbf{100.0} \\
& MSE                  & 0.088 & 0.086 & 0.085 & 0.089 & 0.086 & \textbf{100.0} \\
\hline
\multirow{2}{*}{[30, 90]} & MAE                  & 0.217 & 0.218 & 0.219 & 0.216 & 0.215 & \textbf{100.0} \\
& MSE                  & 0.086 & 0.088 & 0.089 & 0.085 & 0.088 & \textbf{100.0} \\
\hline
\end{tabular}
\end{table}

Table~\ref{tab:percentile} presents an evaluation of performance across a variety of percentile pairs used in the proposed GLID method on different attack methods. The configuration of the percentile pair guides the GLID method in identifying and eliminating outliers. For example, specifying a percentile pair of [10, 70] means that values below the 10$^{\rm th}$ percentile and above the 70$^{\rm th}$ percentile are trimmed away, focusing the analysis on the data within these bounds.
It is observed that, when the percentile pair is set at [10, 70], most methods, except for Zheng attack, register a metric over 100.0, suggesting the models are fully attacked. Similarly, the percentile pair of [10, 90] yields a value over 100 for all methods except Zheng attack. The Zheng attack consistently records low metrics across all settings, such as 0.710, and 0.279 for the pair [10, 70], raising questions about its attack efficacy. On the other hand, FTI shows varied performance; it achieves over 100.0 for most percentile pairs like [10, 70] and [20, 90] but drops to 72.382 and 27.246 for the pair [30, 70].
These results underscore the importance of fine-tuning the percentile pair parameters in the GLID method. Proper parameter selection can effectively trim outliers without significantly impacting overall network performance.

\vspace{-0.2cm}
\section{Conclusion}
In this study, we introduced a novel approach to perform model poisoning attacks on WTP through fake traffic injection. Operating under the assumption that real-world BSs are challenging to attack, we inject fake BS traffic distribution with minimum knowledge that disseminates malicious model parameters. 
Furthermore, we presented an innovative global-local inconsistency detection mechanism, designed to safeguard FL-based WTP systems. 
It employs an adaptive trimming strategy, relying on percentile estimations that preserve accurate model parameters while effectively removing outliers. 
Extensive evaluations demonstrate the effectiveness of our attack and defense, outperforming existing baselines.

\section*{Acknowledgment}
This research was supported by the National Science Foundation through Award CNS--2312138.

\bibliographystyle{IEEEtran}
\bibliography{refs}

\end{document}